# A Bell Theorem Without Inequalities for Two Particles, Using Inefficient Detectors

by

Daniel M. Greenberger
*City College of New York, New York, NY 10031*

Michael Horne
*Stonehill College, Easton, MA 02357*

Anton Zeilinger
*Institute for Experimental Physics, U. of Vienna, Vienna A-1090*

and

Marek Zukowski
*Institute for Theoretical Physics and Astrophysics, U. of Gdansk, 80-952 Gdansk, Poland*

## Abstract

We again consider (as in a companion paper) an entangled two-particle state that is produced from two independent down-conversion sources by the process of "entanglement-swapping", so that the particles have never met. We show that there is a natural extension of the Einstein-Pololsky-Rosen discussion of "elements of reality" to include inefficient detectors. We consider inefficient deterministic, local, realistic models of quantum theory that are "robust", which we consider to be the minimum requirement for them to be taken seriously. By robust, we mean they satisfy the following three criteria: (a) they reproduce the quantum results for perfect correlations, if all particles are detected; (b) they produce some counts for every setting of the angles (so they don't describe some experiments that can easily be performed as "impossible"); (c) all their hidden variables are relevant (they must each produce a detectable result in some experiment). For such models we prove a Greenberger-Horne-Zeilinger (GHZ) type theorem for arbitrary detection efficiencies, showing that any such theory is inconsistent with the quantum mechanical perfect correlations. This theorem holds for individual events with no inequalities.

As a result, the theorem is also independent of any random sampling hypothesis, and we take it as a refutation of such realistic theories, free of the detection efficiency and random sampling "loopholes". The hidden variable analysis depends crucially on the use of two independent laser sources for the down-conversions. We also investigate the necessity of using two independent sources *vs* a single source for all particles. Finally, we argue that the state we use can legitimately be considered as a two-particle state, and used as such in experiments.

# 1. Introduction

We recently produced a Bell's Theorem[1] (in a companion paper, which we refer to as paper A) for two entangled particles that uses a Greenberger-Horne-Zeilinger (GHZ)-type argument[2]. The argument applies to the case where the two particles have a perfect correlation, meaning that if one knows the outcome of a measurement on one of them, one can predict the outcome of a corresponding measurement on the other with absolute certainty, so that an Einstein-Podolsky-Rosen (EPR) element of reality[3] exists. Another feature of the argument is that it involves no inequalities, and discusses only perfectly correlated states.

This argument used a two-particle entangled state that was produced by the method of "entanglement-swapping".[4] In this method, two pairs of particles, each pair in a singlet state, are independently produced. Then one catches one particle of each pair simultaneously (which correlates them into what we call a "cross-entangled" state). This automatically correlates the other particles, which have never met, into an entangled state, the "entanglement-swapped" state. Because the particles have never met and have no shared history, there are many restrictions present that limit the capacity of a deterministic, realistic, local theory to model the behavior of such a state.

Our argument in paper A used counters of 100% efficiency, so it had no need to exploit all the limitations inherent in the system. However there is a natural extension of the idea of reality proposed by EPR that applies to inefficient detectors, and that applies to the type of experiment we are considering. Then, exploiting the EPR locality assumptions, we can prove that the Bell functions that describe the outcome of our experiments for perfect correlations in local, deterministic, realistic models can be factored in such a way that the instructions to the system contained in the hidden variables cannot make use of the angular settings of the polarization rotators used in the experiment. It follows from this that the predictions of such local realistic models are self-contradictory, a result that is true independently of the efficiency of the detectors, for a class of models of inefficient detectors that we call "robust". Models that are not robust are too inefficient to effectively model the experiment, and we do not take them seriously. Within this limitation, this is a new type of result, that can be used to rule out such realistic theories, even when using detectors of low efficiency. We also do not need to assume any kind of random sampling hypothesis, and thus our result closes two of the important loopholes in this field.[5]

We believe that arguments concerning the efficiency of quantum detectors are more substantial than most other classical arguments that attempt to reproduce the quantum results with realistic, local theories, because of the limited efficiencies of actual quantum detectors, especially those involving photons, and one should be able to face and refute such arguments. The experiment we discuss uses the technology of experiments that have already been performed, and the Zeilinger group is actively planning to perform an experiment using two independent laser sources.

Others have produced arguments very similar to ours, in a different context (see for example the very cleanly written papers of Hardy[6,7], Cabello[8,9], Aravind[10], and Chen[11], and Pavicic and Summhammer[12]). However as explained in paper A, these papers do not discuss in detail individual hidden variable models. Our paper shows the inconsistency of such models for individual values of the hidden variables, since they cannot reproduce the quantum perfect correlation results for all angles.

A recent paper by Broadbent and Méthot[13] argues that entanglement swapping experiments can be explained by local hidden variables. But it gives an example that is much simpler than our experiment, and their results do not apply to our experiment.[14]

Some people would argue that we (and refs. (6-12)) do not have a true two-particle state since we start with a four-particle state and reduce it by subsequent measurements. It is true that one needs all four particles to prove the existence of the various elements of reality present in the two-particle state. But once this is done one can perform and analyze EPR experiments with this two-particle state, and it yields results much stronger than the usual Bell theorems of standard single source two-particle states

We proceed by showing that one can extend the EPR analysis to the case of inefficient detectors. We are analyzing inefficient models that claim to classically reproduce quantum results, but they have to be what we consider to be reasonable models. So to proceed we make three assumptions that are consistent with the quantum results and that restrict the models considered to a class we call "robust" models, a restriction we consider to be reasonable for any theory that tries to mimic the quantum results, even inefficiently.

First, we assume that if all four particles are detected, the result will agree with the quantum mechanical predictions for perfect correlations. The second assumption concerns the number of counts detected in a given experiment. Quantum mechanics predicts that in a given experiment the Bell states appearing for the central two particles, and those for the two outer particles, are correlated. If the two central particles are in either of the Bell states $\phi_{bc}^+$ or $\psi_{bc}^-$ (see eq. (4) of paper A), then the two outer particles will also be in one of these two states. The total number of events detected in one or the other of these two states, $N_+$, will then be a fixed number that depends on the efficiency of the counters, but that will be independent of the angular settings for each particle. (A similar result holds for the other two Bell states, yielding $N_-$ events.) For perfect detectors, $N_+ = N_- = \frac{1}{2}N_0$, where $N_0$ is the number of possible events. For inefficient counters, $N_+ = N_- = \frac{1}{2}\eta N_0$, where $\eta$ is the combined efficiency of the detectors, independently of the angles involved. This is what quantum theory predicts, but our second assumption is much weaker than this quantum result, and merely requires that both $N_+$ and $N_-$ are $\neq 0$, for any settings of the angles, $\varphi_i$. Otherwise, there will be angles for which it is impossible to have any measurable events at all. It is not possible to prove that a theory that produces no events at all is inconsistent. But if they do produce some events at every angular setting, then they will be inconsistent!

The third assumption is that all the hidden variables are relevant, meaning that they must each contribute to some experiment $N[\varphi_i]$. If they do not, they have no operational significance whatsoever, and we have no means to verify their presence.

We call models that fall within these restrictions "robust", and the proof will hold for such models. Without these limitations, one can make models that are so extremely inefficient that, e. g., they can agree with quantum mechanics for one or two measurement angles, and then declare that the detectors will never fire in any other situation. Such a model does agree with quantum mechanics where it works, and so it is consistent, but it almost never works! So some restrictions on inefficiency are inevitable, and we consider robustness to be reasonable. The conditions for robustness are spelled out in eqs. (5), (15), and (16).

From these three assumptions (robustness) we can prove that each of the EPR functions can be factored into a product of two terms, one depending on the angular setting, and one depending on the hidden variables involved. This factorization enables one to prove that the entire EPR scheme is inconsistent, as in the case for efficient detectors. Because the theories based on the scheme are internally inconsistent one does not need an experiment to rule them out. They are self-defeating. One only has to show that the quantum mechanical perfect correlations are correct. In this paper we start with the assumption that quantum mechanics works, and prove that robust, local realistic, deterministic theories do not work.

A priori, the possibility of factorizing these models, which is crucial to our proof, is neither obvious nor intuitive. We shall make it more plausible by first providing a set of consistency conditions that are necessary if a particular model is to be factorized, and then showing that these conditions in fact hold quantum mechanically. Once we see that these necessary consistency conditions are true, we can show that they, together with robustness, are also sufficient to prove factorizability of the models.

We note that we are assuming from the start that one has a consistent mathematical model that is local, realistic, and deterministic, and that can explain all the perfect correlations, eq. (1), in the experiment we are analyzing, so that for these correlations it agrees with quantum theory, and it prescribes all the functions, $A(\varphi_1,\lambda_1), D(\varphi_4,\lambda_4),$ and $F(\varphi_2,\varphi_3,\lambda_1,\lambda_4),$ that are necessary to do this. One cannot obtain these functions $A, D,$ and $F$, from experiment, since the $\lambda$ are by definition "hidden". Our goal is to prove that even assuming such a model exists, if it is robust we can show a contradiction—proving that any such model is inconsistent. As a step toward this contradiction, we will prove that if there is a model that can assign a consistent set of $A(\varphi_1,\lambda_1),$ *etc.*, that satisfies eq. (1), then one can also find a model that is equivalent to it (*i.e.*, that assigns the same values to $A(\varphi_1,\lambda_1),$ *etc.*) but at the same time is also factorized, in the sense of eqs. (18), so that this factorization is a property of the original model, even if it is very obscure in the original model.

We assume the results and notation of paper A, and any equations that we use from that paper will be denoted by an A after the equation number (e.g., eq. (4A)). We reproduce here for convenience Fig. (1) from that paper, to refer to the experiment we are describing.

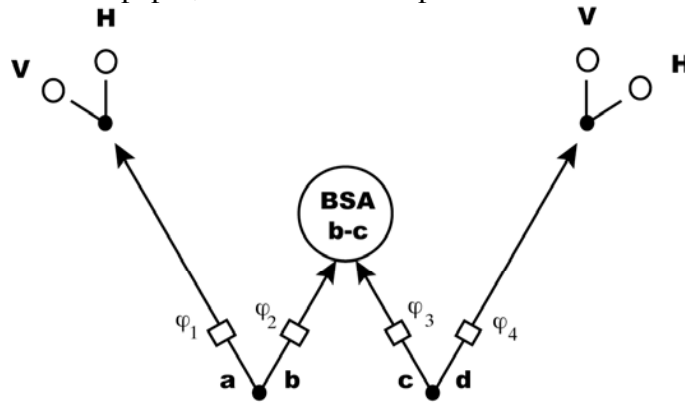

**Fig. (1)**
Figure (1). <u>Schematic Diagram of the Creation of the Two-Particle State</u>
In this experiment there are two independent down-conversions, one creating the pair of photons *a-b*, and the other the pair *c-d*. Each of them undergoes a rotation through the angle $\varphi_i$, and particles *b* and *c* enter a Bell-state-analyzer (BSA), which will annihilate them while detecting which Bell state they were in. If the angles $\varphi_i$ are set properly, as one of the perfect correlation cases, this process forces the particles a and d into a two-particle Bell state. In the actual experiment the Bell state of *a* and *d* is not determined, only their polarizations, but this is sufficient to rule out locally realistic, deterministic theories as an explanation of their observed properties.

We also reproduce eq. (14A), which describes all the perfect correlations in this experiment, which must be described by any classical, deterministic, realistic, local description of the experiment,

$$A(\varphi_1, \lambda_1) F_{\kappa(\lambda_1, \lambda_4)}(\varphi_2, \varphi_3, \lambda_1, \lambda_4) D(\varphi_4, \lambda_4) = 1, \quad \zeta_\kappa = 0, \pm \pi,$$

$$A(\varphi_1, \lambda_1) F_{\kappa(\lambda_1, \lambda_4)}(\varphi_2, \varphi_3, \lambda_1, \lambda_4) D(\varphi_4, \lambda_4) = -1, \quad \zeta_\kappa = \pm \frac{\pi}{2}, \quad (1)$$

$$\zeta_\kappa = \varphi_1 - \varphi_2 + \kappa(\lambda_1, \lambda_4)(\varphi_3 - \varphi_4), \quad (\zeta_+ = \xi, \ \zeta_- = \eta).$$

Eq. (1) records the product of the polarizations of the four particles.

## 2.  Extending the EPR Analysis to the Case of Inefficient Detectors

These results hold in the case where the detectors are 100% efficient, which means that the functions $A(\varphi_1, \lambda_1)$, $D(\varphi_4, \lambda_4)$, $F_\kappa(\varphi_2, \varphi_3, \lambda_1, \lambda_4)$, and $\kappa(\lambda_1, \lambda_4)$, exist according to the EPR postulates, and are equal to $\pm 1$ for every value of their arguments, which in turn means that every one of the four photons that is generated in each event is counted at a detector. In paper A we showed that this situation, given by eq. (1), is inconsistent. Now we shall assume that this 100% efficiency is not necessarily the case, but rather that the particles may reach their detectors and not be counted.

This introduces a complication into the argument since the existence of the functions *A, D, κ,* and *F,* depends critically on the EPR postulates. However if the particle is not always counted, then one no longer has the one-to-one correspondence between predictability and reality needed to define an element of reality, and therefore completeness. Nonetheless, if we are considering a realistic, deterministic model, there is a natural extension of the EPR argument to cover this case.

In any experiment where the conditions for a perfect correlation are met, namely where $\zeta_\kappa = 0, \pm \frac{\pi}{2}, \pm \pi$, if we successfully detect three of the particles in a given event, there are only two possibilities for the fourth particle. The first is that we detect it, in which case we can predict in advance what its polarization will be. If this happens to be particle *a*, say, we can say that in this case $A(\varphi_1, \lambda_1)$ exists, and has the value $\pm 1$, which was determined when the particle was created. The second possibility is that it passes through its detector, but is not detected. (That it does pass through the detector is a consequence of energy and momentum conservation, and is actually an element of reality.) But because it is not detected, it has no further effect on the experiment, and we can consistently assign to $A(\varphi_1, \lambda_1)$ the value $A = 0$.

In a deterministic theory, we can assume that this value was assigned to the particle when it was created. In other words this photon, with these particular values of $\lambda_1$ and $\varphi_1$, was destined at its creation not to be detected. The alternative is that the particle is not recorded simply because the detector is inefficient. It counts only a certain percentage of particles impinging upon it, independently of any state variables $\lambda_i$, and angles $\varphi_i$, that may determine the properties of the particles. This case is conceptually rather simple in that one may then merely consider those particles that are counted, knowing that one is counting a fair sampling of all the particles that impinge upon the counters. Then the outcome is independent of the properties of the interaction of the counters with the particles, except for random efficiency effects which will not prejudice any results one might obtain in the case of 100% efficiency and one can apply Bell-like theorems in this case. We will not be concerned with this case in what follows. We are concerned with a deterministic theory, for which no random sampling assumptions need be made. This case is more general than the stochastic case mentioned above, since a deterministic theory can be modelled to duplicate the results of a stochastic theory.

One may well question whether what we have left after extending the EPR theory to inefficient counters can truly be called an "element of reality". The answer is definitely "yes", because one must remember the motivation for introducing the term. Since after making a measurement, one can predict a property of the particle without in any way interacting with it, then according to EPR we cannot have affected this property, and so the property must have existed before we made the measurement. Thus this is a true, objective property of the particle that it must have possessed since it was created, or at least since it last interacted with another particle, and hence the designation "element of reality". This argument still holds in our situation since, while we cannot predict whether it will be detected, we can predict this property precisely, *if* it is detected. Thus the particle must either possess this property beforehand, or it must be determined beforehand that it will not be detected. In either case, the existence of the property does not depend on the measurement, and so it is an objective element of reality.

Everything we have said about particle *a* also applies to particle *d*. So the functions *A* and *D* are to be considered as deterministic functions representing instructions to the particle not only to have a particular polarization if it is counted, but also to determine whether the particle is to be counted or not. Specifically, we will amend the definitions of the functions *A* and *D* in the inefficient case to read

$$A(\varphi_1, \lambda_1) = \pm 1, 0; \quad D(\varphi_4, \lambda_4) = \pm 1, 0, \tag{2}$$

In eq. (2), no limits are placed on the functions, except that we will demand the consistency condition that the product of all the functions agrees with the quantum theory results for perfect correlations whenever all four particles are actually detected. The existence of these functions extends the concept of completeness to the case of inefficient counters.

The situation for particles *b* and *c* is similar, but a little more subtle. These particles are not counted separately, but as part of an entangled state. In our experiment, particles *a* and *d* are individually counted, and so we do not learn the value of $\kappa$. However, we do not have to run our particular experiment. We could have instead combined particles *a* and *d* at a Bell-state analyzer (BSA), measuring their Bell state. (See the experiment depicted in Fig. (2) of paper A.) Such an experiment would reveal the value of the functions $\kappa$ and $G_\kappa$, the equivalent of $F_\kappa$, but for particles *a* and *d*. Then the value of the function $F_\kappa$ would be known. Since the particles *b* and *c* have no information as to whether particles *a* and *d* will be detected separately, or combined into a Bell state, we must assume that the latter possibility is taken into account in the function $F_\kappa$, so that the function $\kappa(\lambda_1, \lambda_4)$ must be assigned at the outset. (The details of this argument are in paper A).

There is a further point to be made concerning particles *b* and *c*. When they meet at the BSA, information becomes available from both hidden variables, $\lambda_1$ and $\lambda_4$, and this information may indicate a possible violation of eq. (4A), and so trigger an included instruction that one or both of the particles should not be counted. So their combined effect can be rather subtle.

Nonetheless, the output of the BSA is a Bell state whose properties obey the EPR criteria for elements of reality when both particles are detected, so the situation for the function $F_\kappa$ becomes similar to that for the other two functions, except that it is determined by both the hidden variables $\lambda_1$ and $\lambda_4$.

As in the case for the functions A and D, if the counters are not 100% efficient, the function F can assume the value 0, when the measurement does not reveal the Bell state of particles $b$ and $c$ (possibly because the particles are not both counted). Therefore, for $F$, as for $A$ and $D$,

$$F_\kappa(\varphi_2, \varphi_3, \lambda_1, \lambda_4) = \pm 1, 0. \tag{3}$$

Here, the $\pm 1$ values represent the product of their polarizations. In all cases the situation at the detectors will have been deterministically decided by a set of instructions set up when the particles were created, but which may be flexible enough to alter the particle's behavior in response to any new information available when both particles arrive at the BSA.

As mentioned earlier, we will also be guided here by a further assumption, that while the counters may be inefficient for various reasons, they will not violate the perfect correlation results of quantum mechanics, when all particles are detected. This has the non-trivial consequence that whether or not all the particles are counted, eq. (1) becomes,

$$A(\varphi_1, \lambda_1) F_\kappa(\varphi_2, \varphi_3, \lambda_1, \lambda_4) D(\varphi_4, \lambda_4) = \begin{cases} 1, \, or \, 0, & \zeta_\kappa = 0, \pm\pi, \\ -1, \, or \, 0, & \zeta_\kappa = \pm\frac{\pi}{2}, \end{cases}$$

$$\kappa(\lambda_1, \lambda_4) = \pm 1. \tag{4}$$

For 100% efficient detectors, this set of equations is the same as eq. (1), which we have shown in paper A to be inconsistent. But if the product can sometimes be 0, then the situation is no longer apparent, and it may or may not be possible to satisfy eq. (4). We shall show in Section 6 that for the class of robust models no matter how inefficient the detectors are, the eqs. (4) are incompatible with any local, realistic theory consistent with our assumptions.

The perfect correlations, eq. (9) only hold for certain values of $\zeta_\kappa$. However, in the factor $A(\varphi_1, \lambda_1)$, both $\varphi_1$ and $\lambda_1$ can be independently varied. And similarly for the parameters $\varphi_4$ and $\lambda_4$ in the factor $D(\varphi_4, \lambda_4)$. This is also true for the parameters in $F(\varphi_2, \varphi_3, \lambda_1, \lambda_4)$. The perfect correlation restrictions, *e.g.* to $\zeta_\kappa = 0$, represent a restriction on the variables in eq. (4). For other values of $\zeta_\kappa$ the right hand side of the equation exists and can take any value ($\pm 1$, 0).

In order to study the effect of detector efficiency more carefully we rewrite eq. (4) as

$$A(\varphi_1, \lambda_1) F_\kappa(\varphi_2, \varphi_3, \lambda_1, \lambda_4) D(\varphi_4, \lambda_4) = \Delta_A(\varphi_1, \lambda_1) \Delta_{F_\kappa}(\varphi_2, \varphi_3, \lambda_1, \lambda_4) \Delta_D(\varphi_4, \lambda_4), \quad \zeta_\kappa = 0, \pm\pi,$$

$$A(\varphi_1, \lambda_1) F_\kappa(\varphi_2, \varphi_3, \lambda_1, \lambda_4) D(\varphi_4, \lambda_4) = -\Delta_A(\varphi_1, \lambda_1) \Delta_{F_\kappa}(\varphi_2, \varphi_3, \lambda_1, \lambda_4) \Delta_D(\varphi_4, \lambda_4), \quad \zeta_\kappa = \pm\frac{\pi}{2}, \quad (5)$$

$$\Delta_A(\varphi_1, \lambda_1) = \begin{cases} 1, & |A| = 1, \\ 0, & |A| = 0, \end{cases} \quad \Delta_{F_\kappa}(\varphi_2, \varphi_3, \lambda_1, \lambda_4) = \begin{cases} 1, & |F_\kappa| = 1, \\ 0, & |F_\kappa| = 0, \end{cases} \quad \Delta_D(\varphi_4, \lambda_4) = \begin{cases} 1, & |D| = 1, \\ 0, & |D| = 0. \end{cases}$$

The functions $\Delta_A$, $\Delta_D$, are non-zero when their respective detectors fire, detecting a particle; the function $\Delta_{F_\kappa}$ is non-zero when the Bell state and polarizations of particles $b$ and $c$ are detected; and the product $\Delta_A \Delta_{F_\kappa} \Delta_D = 1$ when all detectors fire, corresponding to the registering of an event. The functions $\Delta$ merely define the range over which the functions $A$, $D$, and $F_+$ are non-zero, and numerically they are merely the absolute values of these functions. They obey the relations

$$\Delta_A = |A|, \quad \Delta_{F_\kappa} = |F_\kappa|, \quad \Delta_D = |D|,$$
$$\Delta_A^2 = \Delta_A, \quad \Delta_{F_\kappa}^2 = \Delta_{F_\kappa}, \quad \Delta_D^2 = \Delta_D,$$
$$A\Delta_A = A, \quad F_\kappa\Delta_{F_\kappa} = F_\kappa, \quad D\Delta_D = D, \qquad (6)$$
$$A^2 = \Delta_A, \quad F_\kappa^2 = \Delta_{F_\kappa}, \quad D^2 = \Delta_D,$$

and again, the functions $A$, $D$, and $F$, can take the values $\pm 1,0$ (see eqs. (2) and (3)), while the $\Delta$'s are restricted to $+1,0$. We shall use these functions in the next section. For notational convenience we define the set $\Omega_\kappa$ by

$$(\lambda_1, \lambda_4) \in \begin{cases} \Omega_+(\varphi_1, \varphi_2, \varphi_3, \varphi_4; \lambda_1, \lambda_4), & \text{if } \kappa(\lambda_1, \lambda_4) = +1, \text{and } \Delta_A\Delta_{F_\kappa}\Delta_D = 1, \\ \Omega_-(\varphi_1, \varphi_2, \varphi_3, \varphi_4; \lambda_1, \lambda_4), & \text{if } \kappa(\lambda_1, \lambda_4) = -1, \text{and } \Delta_A\Delta_{F_\kappa}\Delta_D = 1. \end{cases} \qquad (7)$$

Thus $\Omega_\kappa$ contains all those points $\Omega_+$, and $\Omega_-$, for which all of the detectors fire, leading to the knowledge that particles $b$ and $c$ are in a definite Bell state, and that particles $a$ and $d$ have a specific polarization. Below we shall work with the case $\kappa = +1$, although a similar argument could be made for the case $\kappa = -1$.

Let $[\varphi_1, \varphi_2, \varphi_3, \varphi_4]$ represent an experiment where the four angles $\varphi_i$ are defined by the brackets. Then for a locally realistic interpretation of the experiment, define $\gamma_{\lambda_1, \lambda_4}[\varphi_i]$ as the outcome of this experiment (namely, $+1$, $0$, or $-1$) for one event, produced by a particular value of the hidden variables $(\lambda_1, \lambda_4)$, so that

$$\gamma_{\lambda_1, \lambda_4}[\varphi_i] = A(\varphi_1, \lambda_1)F_+(\varphi_2, \varphi_3, \lambda_1, \lambda_4)D(\varphi_4, \lambda_4), \qquad (8)$$

regardless of whether $\zeta_+ = 0$, or not. Then the number of events that will be recorded by all the detectors, *i.e.*, events in $\Omega_+$, will be

$$N_+[\varphi_i] = \tfrac{1}{2} N_0 \iint_{\lambda_1, \lambda_4 \in \Omega_+} d\lambda_1 d\lambda_4 \rho_1(\lambda_1) \rho_4(\lambda_4) \left| \gamma_{\lambda_1, \lambda_4}[\varphi_i] \right|, \qquad (9)$$

where $N_0$ is the total number of events generated, that would be counted if all the detectors were 100% efficient, and the $\rho_i(\lambda_i)$ are positive semi-definite weighting functions, such that $\int d\lambda_i \rho_i(\lambda_i) = 1$. The $\tfrac{1}{2}$ comes from the fact that only half the total counts are involved, namely those with $\kappa = +1$. (We note that the integral defining $N_+[\varphi_i]$ does not factor into separate integrals over $\lambda_1$ and $\lambda_4$, because presumably, $\Omega_\kappa(\lambda_1, \lambda_4)$ does not.)

Quantum theory predicts that $N_+[\varphi_i] = \tfrac{1}{2} \eta N_0$, depending on the overall efficiency $\eta$ of the detectors, which should be the product of the efficiencies of the individual detectors, and it does not depend on the $\varphi_i$. Explicitly, we have from eq. (4A) for the quantum case, for photons,

$$P(H_a, H_d, \phi_{bc}^+) = P(V_a, V_d, \phi_{bc}^+)$$
$$= P(H_a, V_d, \psi_{bc}^-) = P(V_a, H_d, \psi_{bc}^-) = \tfrac{1}{8}\cos^2\zeta_+,$$
$$P(H_a, V_d, \phi_{bc}^+) = P(V_a, H_d, \phi_{bc}^+)$$
$$= P(H_a, H_d, \psi_{bc}^-) = P(V_a, V_d, \psi_{bc}^-) = \tfrac{1}{8}\sin^2\zeta_+, \qquad (10)$$

where $P(H_a, V_d, \psi_{bc}^-)$ means the probability for finding that particle $a$ has a horizontal, and particle $d$ a vertical polarization, and particles $b$ and $c$ are in the Bell state $\psi_{bc}^-$, etc. The sum of these 8 probabilities is only $\tfrac{1}{2}$, because eq. (10) only represents the case where we have $\phi_{bc}^+$ *and* $\psi_{bc}^-$, (corresponding to the $\kappa = +1$ case, for a local, realistic theory). (Also note that in the case of spin ½ particles, in eq. (10) and below in eq. (11), read ↑ for H, and ↓ for V.) In the quantum case if (as in the original experiment) we only count states for which the BSA gives $\psi_{bc}^-$, the singlet state,

$$
\begin{aligned}
N_+(\psi_{bc}^-, \zeta_+) &= \eta N_0 \Big[ (P(H_a, V_d, \psi_{bc}^-) + P(V_a, H_d, \psi_{bc}^-)) \\
&\quad + (P(H_a, H_d, \psi_{bc}^-) + P(V_a, V_d, \psi_{bc}^-)) \Big] = \tfrac{1}{4}\eta N_0, \\
E_{+,qu}(\psi_{bc}^-, \zeta_+) &= \frac{\eta N_0}{N_+(\psi_{bc}^-, \zeta_+)} \Big[ (P(H_a, H_d, \psi_{bc}^-) + P(V_a, V_d, \psi_{bc}^-)) \\
&\quad - (P(H_a, V_d, \psi_{bc}^-) + P(V_a, H_d, \psi_{bc}^-)) \Big] = -\cos 2\zeta_+,
\end{aligned}
\tag{11}
$$

where $E_{+,qu}(\psi_{bc}^-, \zeta_+)$ is the quantum expectation value of the product of the polarizations of the four particles, given that the middle two are in the singlet state. In the classically realistic case, this would be

$$
E_{+,class.}[\varphi_i] = \frac{\iint_{\varphi_i, \lambda_1, \lambda_4 \in \Omega_+, F_+ = -1} d\lambda_1 d\lambda_4 \rho_1(\lambda_1)\rho_4(\lambda_4)\gamma[\varphi_i]}{\iint_{\varphi_i, \lambda_1, \lambda_4 \in \Omega_+, F_+ = -1} d\lambda_1 d\lambda_4 \rho_1(\lambda_1)\rho_4(\lambda_4)|\gamma[\varphi_i|}, \tag{12}
$$

where the restriction to $F_+ = -1$ is to the singlet state, $\psi^-$ (see eq. (10A)).

We are trying to show that deterministic models cannot reproduce the results of quantum mechanics. In order to make that possible, we have required three overall conditions, which are satisfied by quantum theory, that should also be satisfied by any candidate model, even before we seriously examine the model. The first of these considerations, as we have mentioned, is that when all four particles are detected, and the result is a perfect correlation, so that quantum mechanics gives a definite result 100% of the time, then the classical model must yield the same result. Otherwise it fails in its basic task, namely to provide an alternate explanation of the quantum result. For the second condition, note that if we examine eq. (4A), when one measures only the cases $F_+$, for the results at the central counters $b$ and $c$, one finds that

$$
N_+ = N[\phi_{bc}^+; \varphi_1, \varphi_2, \varphi_3, \varphi_4] + N[\psi_{bc}^-; \varphi_1, \varphi_2, \varphi_3, \varphi_4] = \tfrac{1}{2}\eta N_0, \tag{13}
$$

for any set of angles $\varphi_i$, where $N_0$ is the total number of events, whether detected or not, and $\eta$ represents the combined efficiency of the four detectors. Thus

$$
\begin{aligned}
N_+[\varphi_1, \varphi_2, \varphi_3, \varphi_4] &\neq 0, \\
&\text{for any value of the } \varphi_i.
\end{aligned}
\tag{14}
$$

We shall either take it as a blatant contradiction of quantum mechanics, sufficient to rule a model out, if we can show that $N_+[\varphi_1, \varphi_2, \varphi_3, \varphi_4] = 0$, for some value of the $\varphi_i$, or equivalently that the model is not worth serious consideration. (Similar results hold for $N_-$.) This is weaker than the quantum condition, eq. (13), as it merely requires that some events take place. It follows as a consequence of eq. (14) that

$$
\begin{aligned}
&\text{for every set } \varphi_i, \text{ there exists some } (\lambda_1, \lambda_4), \text{ such that} \\
&\Delta_A(\varphi_1, \lambda_1)\Delta_{F_+}(\varphi_2, \varphi_3, \lambda_1, \lambda_4)\Delta_D(\varphi_4, \lambda_4) \neq 0.
\end{aligned}
\tag{15}
$$

Otherwise, $N_+[\varphi_1, \varphi_2, \varphi_3, \varphi_4] = 0$, in violation of eq. (14).

Thirdly, we shall also impose one further operational condition on the $\lambda$'s, namely that they be relevant to the calculation of the $N_+$. By this we mean that for every $\lambda_1$, there exists some value of $\lambda_4$, and some set $\varphi_i$, such that

$$\Delta_A(\varphi_1, \lambda_1) \Delta_{F_\pm}(\varphi_2, \varphi_3, \lambda_1, \lambda_4) \Delta_D(\varphi_4, \lambda_4) \neq 0;$$

for every $\lambda_4$, there exists some value of $\lambda_1$, and some set $\varphi_i$, such that (16)

$$\Delta_A(\varphi_1, \lambda_1) \Delta_{F_\pm}(\varphi_2, \varphi_3, \lambda_1, \lambda_4) \Delta_D(\varphi_4, \lambda_4) \neq 0.$$

If eq. (16) were not true, then for that value of $\lambda_1$ or that value of $\lambda_4$, there would be no events $N_+[\varphi_i]$, for any set of angles, and so that value of $\lambda_i$ would never produce a count in any experiment. One could then proliferate values of $\lambda_i$ whose only effect would be to lower the overall efficiency of all experiments, and since their presence could never be detected, they would have no operational significance.

These three conditions, eqs. (5), (15), and (16), are sufficient to prove that the functions $A$, $F$, and $D$, can each be factored. We consider these three conditions, viz., that the model does not disagree with quantum theory when all the detectors count, that for every set of angles there are some counts, and that all the $\lambda$'s are relevant, i.e., that they lead to a count at some angle, to be a reasonable minimal requirement for any deterministic model. We call any model that satisfies these conditions robust. Our proof will hold for such robust models, and we will show that they are inconsistent, *i.e.*, they cannot do what they were set up to do, namely reproduce the perfect correlations of quantum theory..

## 3. Factorization of the Functions

We will show that for robust models the functions *A, F,* and *D* in eqs. (5) each can be factored into a product of a function of their angles times a function of their hidden variables. This factorization specifically depends on there being two independent sources, meaning that the two original down-conversions were created by independent lasers, so that the hidden variables, $\lambda_1$ and $\lambda_4$, are truly independent. (This statement has no meaning within quantum theory.) It is valid regardless of whether one has efficient or inefficient detectors. This result is in fact the central theorem of the paper.

It will be convenient to write the functions in the following form. We can write

$$A(\alpha, \lambda_1) = a(\alpha, \lambda_1) \Delta_A(\alpha, \lambda_1),$$

$$D(\alpha, \lambda_4) = d(\alpha, \lambda_4) \Delta_D(\alpha, \lambda_4),$$

$$F_\pm(\alpha, \beta, \lambda_1, \lambda_4) = f_\pm(\alpha, \beta, \lambda_1, \lambda_4) \Delta_{F_\pm}(\alpha, \beta, \lambda_1, \lambda_4), \qquad (17)$$

$$a(\alpha, \lambda_1) = \pm 1, \quad d(\alpha, \lambda_4) = \pm 1, \quad f_\kappa(\alpha, \beta, \lambda_1, \lambda_4) = \pm 1,$$

$$\Delta_A = 1, 0, \quad \Delta_{F_\pm} = 1, 0, \quad \Delta_D = 1, 0,$$

where the $a(\alpha, \lambda_1), d(\alpha, \lambda_4),$ and $f_\kappa(\alpha, \beta, \lambda_1, \lambda_4)$, represent the value of $A$, $D$, and $F_\kappa$, while the $\Delta$'s define their range, i.e., whether they equal 0 or not. If any of the $\Delta$'s equal 0, the corresponding functions $a, f_\kappa,$ and $d$ are for the moment ambiguous.

Our factorization will take the form

$$A(\alpha, \lambda_1) = a(\alpha)u(\lambda_1)\Delta_A(\alpha, \lambda_1),$$
$$D(\beta, \lambda_4) = a(\beta)v(\lambda_4)\Delta_D(\alpha, \lambda_4),$$
$$F_\kappa(\alpha, \beta, \lambda_1, \lambda_4) = a(\alpha)a(\beta)u(\lambda_1)v(\lambda_4)\Delta_{F_\kappa}(\alpha, \beta, \lambda_1, \lambda_4), \tag{18}$$

where $a, u$ and $v, = \pm 1$, while $\Delta = 0, 1$.

If the function $A(\alpha, \lambda_1)$ is to be factorizable as in eq. (18), then there are two key consistency relations that must be satisfied, which are suggested by the form of eq. (18). The first is

if $A(\alpha, \lambda_1) = a(\alpha)u(\lambda_1)$,  and $D(\beta, \lambda_4) = a(\beta)v(\lambda_4)$, then

$$A(\alpha, \lambda_1)D(\beta, \lambda_4) = a(\alpha)u(\lambda_1)a(\beta)v(\lambda_4) = A(\beta, \lambda_1)D(\alpha, \lambda_4), \text{ or } 0,$$
$$A(\alpha, \lambda_1)A(\beta, \lambda_1)D(\alpha, \lambda_4)D(\beta, \lambda_4) = 1, 0. \tag{19}$$

The second consistency relation is

if $A(\alpha, \lambda_1) = a(\alpha)u(\lambda_1)$,  then

$$A(\alpha, \lambda_1)A(\beta, \lambda_1') = a(\alpha)u(\lambda_1)a(\beta)u(\lambda_1') = A(\alpha, \lambda_1')A(\beta, \lambda_1), \text{ or } 0;$$
$$A(\alpha, \lambda_1)A(\alpha, \lambda_1')A(\beta, \lambda_1)A(\beta, \lambda_1') = 1, 0. \tag{20}$$

and similarly for $D(\alpha, \lambda_4)$.

The actual perfect correlations, eq. (4), implies that these relations are satisfied. (We shall restrict ourselves to the case $\kappa = +1$. There is an equivalent proof for the case $\kappa = -1$.) Consider, whenever there exists an event

$$A(\alpha, \lambda_1)F_+(\gamma, \gamma, \lambda_1\lambda_4)D(\alpha, \lambda_4) = 1, \quad \zeta_+ = 0. \tag{21}$$

If there exists another event,

$$A(\beta, \lambda_1)F_+(\gamma, \gamma, \lambda_1\lambda_4)D(\beta, \lambda_4) = 1, \quad \zeta_+ = 0, \tag{22}$$

for the same $\gamma, \lambda_1$, and $\lambda_4$, then by multiplying eqs. (21) and (22), we get eq. (19). If eq. (19) holds for two different values of $\lambda_1$,

$$A(\alpha, \lambda_1)A(\beta, \lambda_1)D(\alpha, \lambda_4)D(\beta, \lambda_4) = 1,$$
$$A(\alpha, \lambda_1')A(\beta, \lambda_1')D(\alpha, \lambda_4)D(\beta, \lambda_4) = 1, \tag{23}$$

then by multiplying both eqs. (23) together, we get eq. (20).
A very similar proof applies to the function $D(\alpha, \lambda_4)$, so that eqs. (19) and (20) also hold for the functions $D$.

The form of eq. (18) also suggests a number of consistency relations for the function $F_+$. We write down a few of them, although many variations of these exist, and any consistency relation suggested by the factorization in eq. (18) must be true in any model, or else the factorization would be inconsistent. For example, the following must all be true, if the functions $\neq 0$ :

$$F_+(\alpha, \beta, \lambda_1, \lambda_4) = F_+(\beta, \alpha, \lambda_1, \lambda_4),$$
$$F_+(\alpha, \beta, \lambda_1, \lambda_4) = F_+(\alpha, \gamma, \lambda_1, \lambda_4')F_+(\beta, \gamma, \lambda_1', \lambda_4')F_+(\delta, \delta, \lambda_1', \lambda_4'),$$
$$F_+(\alpha, \beta, \lambda_1, \lambda_4)F_+(\alpha, \gamma, \lambda_1, \lambda_4') = F_+(\delta, \beta, \lambda_1', \lambda_4)F_+(\delta, \gamma, \lambda_1', \lambda_4'),$$
$$F_+(\gamma, \gamma, \lambda_1, \lambda_4)F_+(\delta, \delta, \lambda_1', \lambda_4') = F_+(\alpha, \beta, \lambda_1, \lambda_4)F_+(\alpha, \beta, \lambda_1', \lambda_4'), \tag{24}$$

etc.

These are representative of the various relations that exist.

We will not bother to prove all of the relations in eq. (24), but point out that all such relations are true. The way to prove them is to insert the appropriate $A$ and $D$ functions, and use eqs. (5), (19), and (20). We will illustrate the procedure with the top eq. of eqs. (24),

$$A(\alpha,\lambda_1)F_+(\alpha,\beta,\lambda_1,\lambda_4)D(\beta,\lambda_4) = 1,$$
$$A(\beta,\lambda_1)F_+(\beta,\alpha,\lambda_1,\lambda_4)D(\alpha,\lambda_4) = 1. \tag{25}$$

Then, multiply the two eqs. (25) together, and use eq. (19) to eliminate all the $A$'s and $D$'s. In this way one proves that all the consistency relations suggested by the factorization in eq. (18) hold, when the functions $\neq 0$.

That eqs. (19), (20), and (24) hold guarantees that when one is able to assign values to the functions $a(\alpha)$, $u(\lambda_1)$, and $v(\lambda_4)$, one will not arrive at contradictions. We emphasize that eqs. (19), (20), and (24), are consistency conditions that the functions $A, D,$ and $F$ must satisfy if they are factorable. They are not a proof of factorizability. However, we can use them to construct a factorization that is necessarily consistent.

One should note at the outset that the factorization will never be unique. Even in the case of 100% efficiency of the detectors, one can change $a(\alpha)$ to $-a(\alpha)$ and $u(\lambda_1)$ to $-u(\lambda_1)$ everywhere, without affecting $A(\alpha,\lambda_1)$, which is just the product of the two functions. As the efficiency of the counters becomes very low, there will be subsets S:$(\alpha, \lambda_1)$ of the angles and hidden variables that do not interact with other subsets S':$(\alpha', \lambda_1')$ (meaning that $A(\alpha,\lambda_1)A(\alpha',\lambda_1') = 0$, for all $(\alpha,\lambda_1) \in$ S, and all $(\alpha',\lambda_1') \in$ S'), and one can change signs within each of the subsets separately. So as the efficiency of the detectors decreases, the non-uniqueness of the assignment becomes greater. Ultimately, for very inefficient counters there will be functions $A(\alpha,\lambda_1)$ for which $A(\alpha,\lambda_1) \neq 0$ for only one value of $\lambda_1$. We are only interested in showing that one can assign a consistent factorization to any model of these Bell functions. The lack of uniqueness is irrelevant, and is in fact guaranteed by the very designation "hidden variables". They can be used to model the behavior of the system, but they are not by their very nature directly accessible to measurement in a one-to one fashion.

There is another extremely important property that both $a(\alpha,\lambda_1)$ and $d(\beta,\lambda_4)$ exhibit, namely that robustness determines that they are defined for all values of $\alpha$, $\lambda_1$, and $\lambda_4$, even though originally it appeared that they were defined only when $\Delta_A$, $\Delta_D$, and $\Delta_F$ were not zero. It follows from eq. (15) that for every value of $\alpha$ there is some value of $\lambda_1$ for which $A(\alpha,\lambda_1) \neq 0$, so that in eq. (18) $a(\alpha)$ is defined for every value of $\alpha$. But for each value of $\alpha$, only some values of $\lambda_1$ and $\lambda_4$ occur for which $A$ and $D \neq 0$. It also follows from eq. (21), that for every value of $\lambda_1$ there is some value of $\alpha$ for which $A(\alpha,\lambda_1) \neq 0$, and so it follows that in eq. (18), $u(\lambda_1)$ is defined for all $\lambda_1$. So robustness extends the definition of $a(\alpha,\lambda_1)$ beyond the region where it was originally defined,(namely where $\Delta_A \neq 0$), to all regions of $(\alpha,\lambda_1)$, and it is never $= 0$. It is only $\Delta_A(\alpha,\lambda_1)$ that can make $A(\alpha,\lambda_1) = 0$ (and similarly for $D$). This will never lead one into trouble however, because of the consistency conditions, and in experiments one only needs regions where $\Delta_A$ and $\Delta_D \neq 0$. These same remarks also follow for the function $F$.

In order to prove that our functions factorize, we start with the equation

$$A(\alpha,\lambda_1)F_+(\beta,\beta,\lambda_1,\lambda_4)D(\alpha,\lambda_4) = 1,0. \tag{26}$$

For a given value of $\alpha$ and $\beta$,the right hand side of eq.(26) must $= 1$ for at least one set of $(\lambda_1, \lambda_4)$, according to eq. (15). If one takes all values of $\alpha$ for which the right hand side $= 1$, one sees that

$$A(\alpha', \lambda_1)D(\alpha', \lambda_4) = A(\alpha, \lambda_1)D(\alpha, \lambda_4), \tag{27}$$

for all such $\alpha'$, so the product is independent of $\alpha$. Therefore, one can take one particular value of $\alpha$, $\alpha_0$, and set

$$A(\alpha_0, \lambda_1)D(\alpha_0, \lambda_4) \equiv w(\lambda_1, \lambda_4) \equiv u(\lambda_1)v(\lambda_4). \tag{28}$$

Since the left hand side factors, the right hand side must also. In a similar way, eq. (26) does not depend on $\beta$. So one may take one value of $\beta$, $\beta_0$, for which $F(\beta, \beta, \lambda_1, \lambda_4) \neq 0$, and one has

$$F_+(\beta_0, \beta_0, \lambda_1, \lambda_4) = u(\lambda_1)v(\lambda_4). \tag{29}$$

One may extend this to other values of $\lambda_1'$ and $\lambda_4'$ by writing

$$\begin{aligned} A(\alpha_0, \lambda_1') &= u(\lambda_1'), \quad \text{for } A(\alpha_0, \lambda_1') \neq 0, \\ D(\alpha_0, \lambda_4') &= v(\lambda_4'), \quad \text{for } D(\alpha_0, \lambda_4') \neq 0. \end{aligned} \tag{30}$$

Finally, one can extend it to different values of $\alpha$ by writing

$$\begin{aligned} A(\alpha, \lambda_1') &= a(\alpha)u(\lambda_1'), \quad \text{for } A(\alpha, \lambda_1') \neq 0, \\ D(\alpha, \lambda_4') &= a(\alpha)v(\lambda_4'), \quad \text{for } D(\alpha, \lambda_4') \neq 0. \end{aligned} \tag{31}$$

Remember that the mathematical model gives $A(\alpha, \lambda_1)$ and $D(\alpha, \lambda_4)$, so one is defining the functions $a$, $u$, and $v$, by this procedure. One extends this to $F$ by using

$$\begin{aligned} A(\alpha, \lambda_1)F_+(\alpha, \beta, \lambda_1, \lambda_4)D(\beta, \lambda_4) &= 1, \quad AF_+D \neq 0, \\ F_+(\alpha, \beta, \lambda_1, \lambda_4) &= a(\alpha)a(\beta)u(\lambda_1)v(\lambda_4). \end{aligned} \tag{32}$$

How far can this procedure take you? It will take you to all the points in the set of points $S_1(\varphi_1, \varphi_2\varphi_3, \varphi_4, \lambda_1, \lambda_4)$ that can be reached starting from our initial point above, by all $A(\alpha, \lambda_1)$, or $D$, or $F \neq 0$. They are reached, for example, by starting from $A(\alpha_0, \lambda_{10}) \neq 0$, and extending it to all $A(\alpha_0, \lambda_{11}) \neq 0$, for a given $\alpha_0$, then to all $\alpha_1$ such that $A(\alpha_1, \lambda_{11}) \neq 0$, compatible with all the $\lambda_{11}$, then to all $\lambda_{12}$ such that $A(\alpha_1, \lambda_{12}) \neq 0$, *etc.*

When the set $S_1$ is exhausted, there will be other sets $S_2$, $S_3$, *etc.*, whose points are all disjoint from each other, such that $S_i \cap S_j = 0$. Each of these sets can be self-consistently set up, independently of the others. Altogether, for robust models, they must span the entire space of all angles and $\lambda$'s. There will be no inconsistencies in assignments because of the consistency relations. But the different subsets $S_i$ are not truly independent of each other. This is because, as we have noted, there is indeed a complete overlap caused by the robustness and factorization.

As an example showing that $a(\alpha, \lambda_1)$ is defined everywhere, even in the region where $\Delta(\alpha, \lambda_1) = 0$, consider the consistency condition, eq.(20), in the case where, say, the last term equals $0$,

$$A(\beta, \lambda_1') = 0, \tag{33}$$

while the first three terms do not. Then

$$\begin{aligned} &A(\alpha, \lambda_1)A(\alpha, \lambda_1')A(\beta, \lambda_1) \\ &= a(\alpha)u(\lambda_1)a(\alpha)u(\lambda_1')a(\beta)u(\lambda_1) \\ &= a(\beta)u(\lambda_1'), \end{aligned} \tag{34}$$

even though $\Delta(\beta, \lambda_1') = 0$. So the perfect correlation condition, eq. (5), and the conditions of eqs. (15) and (16), together are sufficiently strong that they define counterfactually what the value of $A(\beta, \lambda_1')$ would be if the detector were to fire, even when it does not fire.

Once again we emphasize that if three of the detectors fire, and $\zeta_+ = 0$, we can predict with 100% certainty what value the fourth detector would have if it fired. So this value is an EPR element of reality, represented by the functions $a$ and $u$. What we cannot predict is whether it will fire, although in a deterministic model that fact will be determined in advance, and that is the role played by the $\Delta$'s. We note once more that the deterministic model itself predicts $A(\alpha, \lambda_1)$, as well as $D$ and $F$. Because of this, it follows that our "independent" sets $S_i$ are not truly independent, and their values for the functions $a$, $u$, and $v$ must be compatible with those of the other regions, $S_j$, $j \neq i$. However, because of the consistency relations, each set, $S_i$, is either totally correct, or totally incorrect, requiring all its signs to be changed..

In order to insure the consistency of the assignments between the sets, $S_i$, we will use a relationship that we will prove in the next section, namely

$$a(\varphi_1)a(\varphi_2)a(\varphi_3)a(\varphi_4) = 1,$$
$$\text{for } \varphi_1 - \varphi_2 + \varphi_3 - \varphi_4 = 0. \tag{35}$$

Then, for example, if $\varphi_4$ comes from the set $S_2$, while the other three $\varphi$'s come from the set $S_1$, one can use this to determine $a(\varphi_4)$. If one has made the wrong choice, one has to reverse all the assignments in the set $S_2$. Once one knows the $a(\varphi_4)$, one can use this to determine whether $u(\lambda_1)$ is correct, from $u(\lambda_1) = a(\varphi_4, \lambda_1)/a((\varphi_4))$, and whether to reverse all the $u(\lambda_1)$. In this way, one arrives at a complete determination of all the $a(\varphi_4)$, $u(\lambda_1)$, and $v(\lambda_4)$, regardless of how inefficient the detectors are.

The functions, $A$, $D$, and $F$, have no explicit dependence on $\zeta_+$, and in fact, the value of $\zeta_+$ is not available locally at any of the detectors, or the BSA, as any of the $\varphi_i$ can be arranged to be suddenly switched, even after the detectors that are not locally in the path of that $\varphi_i$ have already fired. So $\zeta_+$ is a non-local variable, whose value is uncovered only after all the $\varphi_i$ are known, and this value then determines whether the perfect correlation equations are valid in that individual situation. (A similar result holds for $\zeta_-$.) But the values of $A$, $D$, and $F$, themselves, are locally determined by their arguments. Their factorization, as given in eqs. (18), is a consequence of the use of two independent sources, represented by the independent hidden variables $\lambda_1$ and $\lambda_4$, in performing the VW experiments, and it holds for all values of $\zeta_+$, not just $\zeta_+ = 0$.

Although we have already noted this for other reasons, It is also a consequence of eq. (14) that in an equation like

$$A(\alpha, \lambda_1) = a(\alpha)u(\lambda_1)\Delta_A(\alpha, \lambda_1), \tag{36}$$

we must have

$$a(\alpha) \neq 0, \tag{37}$$

or else any experiment of the form

$$N[\alpha, \varphi_2, \varphi_3, \varphi_4] = 0, \quad \text{for arbitrary } \varphi_i. \tag{38}$$

We can also conclude that

$$u(\lambda_1) \neq 0, \quad v(\lambda_4) \neq 0, \tag{39}$$

or else any equation involving $u(\lambda_1)$ will never contribute to any 4-particle event, or for that matter, any event at all, and thus that particular value of $\lambda_1$ will be totally irrelevant, and can be dropped out of the range of the $\lambda$'s. Thus the restriction due to the inefficiency of the detectors shows up only in the $\Delta$ functions.

Finally, we note that there is an issue of internal consistency, that comes up in two situations, concerning the existence of the functions *a, d,* and *f.* The first such situation is that if the detectors are extremely inefficient, we might find that $A(\alpha,\lambda_1) \neq 0$, but the product $FD = 0$ in all cases. That is, we might have

$$A(\alpha,\lambda_1) \neq 0, \text{ but}$$
$$A(\alpha,\lambda_1)F(\alpha,\beta,\lambda_1,\lambda_4)D(\beta,\lambda_4) = 0,$$
$$F(\alpha,\beta,\lambda_1,\lambda_4)D(\beta,\lambda_4) = 0,$$
$$\text{for all } \beta,\lambda_4. \tag{40}$$

In this case we might never have all the detectors except one firing, and we would never be in the situation where we could make a prediction of the value of any that might fire. Therefore the possibility arises for the model to have either

$$A(\alpha,\lambda_1) = +1, \text{ or } A(\alpha,\lambda_1) = -1, \tag{41}$$

in eq. (26), so that one could never guarantee a perfect correlation for $A(\alpha,\lambda_1)$ in any situation, which would certainly violate our consistency conditions. However this ambiguity cannot occur.

The reason for this is that the situation in eq. (26) cannot come about. For if the product $AFD = 0$, while $A \neq 0$, and if either $F$ or $D \neq 0$, then two of the functions would $\neq 0$, and we could predict the value of the third one, which must be consistent with eq. (10), and so its value would then be unique, and our consistency condition would hold. The only way to challenge this uniqueness would be if eq. (31) held, and both

$$F(\alpha,\beta,\lambda_1,\lambda_4) = 0, \text{ and } D(\beta,\lambda_4) = 0,$$
$$\text{for all } \beta,\lambda_4. \tag{42}$$

But if eq. (42) were true, then $D(\beta,\lambda_4)$ would never contribute to any experiment of the form $N_+[\varphi_1,\varphi_2,\beta,\beta]$, for any $\lambda_4$, violating the condition of eq. (14). So the ambiguity implied by eq. (40) never comes about. It is certainly possible for only one detector to fire (or none of them), but not for all angles, as in eq. (42). (Thus there is a limit as to how bad the detectors can be. Otherwise, they can be so grossly inefficient that they violate our conditions for robustness, eqs. (5), (14), (15), and (16).) We have not explored the question of how inefficient a model can possibly be, and still be robust.

The second situation in which it looks like an ambiguity can occur that would destroy our consistency conditions for factorizability, is in the case where, say, $A = 0$, while the product $FD \neq 0$. Then, it would seem that the product $FD$ could be $\pm 1$, and either would seem to be possible. But here too, the choice is actually restricted. The problem is

$A(\alpha,\lambda_1) = 0$, but $F(\alpha,\beta,\lambda_1,\lambda_4)D(\beta,\lambda_4) \neq 0$.

Is $F(\alpha,\beta',\lambda_1,\lambda_4')D(\beta',\lambda_4') = F(\alpha,\beta,\lambda_1,\lambda_4)D(\beta,\lambda_4),$ \hfill (43)

for any $\beta',\lambda_4'$ for which $F(\alpha,\beta',\lambda_1,\lambda_4')D(\beta',\lambda_4') \neq 0$?

In other words, is the product $FD$ ambiguous, or does it always assume the same value for any case where it is $\neq 0$, even though the relevant $A = 0$? We note that if $A \neq 0$, then

$$A(\alpha,\lambda_1)F(\alpha,\beta',\lambda_1,\lambda_4')D(\beta',\lambda_4') = 1,$$
$$\text{if } A \neq 0, \text{ and } FD \neq 0. \tag{44}$$

So the ambiguity only occurs when $A = 0$. But in fact, the reality conditions determines that

$$F(\alpha,\beta,\lambda_1,\lambda_4)D(\beta,\lambda_4) = a(\alpha)a(\beta)u(\lambda_1)v(\lambda_4)a(\beta)v(\lambda_4)$$
$$= a(\alpha)u(\lambda_1), \tag{45}$$

even if $A(\alpha,\lambda_1) = 0$, as in eq. (34). Then eq. (44) will obviously be true, so whenever the product $FD \neq 0$, it must always have the same value, which must be consistent with the case when $A(\alpha,\lambda_1) \neq 0$.

### 4. The Inconsistency of Inefficient Realistic Deterministic Models

Once we have proven that the various functions that are introduced by the EPR reality argument are factorable, one can show straightforwardly that the scheme is inconsistent, even for very inefficient detectors. An important thing to note in what follows is that for any set of angles that yields a perfect correlation, say $\zeta_+ = 0$, for which all detectors fire,

$$A(\varphi_1,\lambda_1)F_+(\varphi_2,\varphi_3,\lambda_1,\lambda_4)D(\varphi_4,\lambda_4) = 1,$$
$$\zeta_+ = \varphi_1 - \varphi_2 + \varphi_3 - \varphi_4 = 0, \tag{46}$$

both $\lambda_1$ and $\lambda_4$ drop out of this equation, i.e., the equation is true for any $\lambda_1$ and $\lambda_4$ consistent with with that choice of the $\varphi_i$. Within the subset of those values of the $\lambda$'s for which eq. (51) holds, the right hand side of the equation is a constant. For any set of angles consistent with

$$A(\varphi_1,\lambda_1)F_+(\varphi_2,\varphi_3,\lambda_1,\lambda_4)D(\varphi_4,\lambda_4) = \Delta_A(\varphi_1,\lambda_1)\Delta_{F_+}(\varphi_2,\varphi_3,\lambda_1,\lambda_4)\Delta_D(\varphi_4,\lambda_4),$$

where $\zeta_+ = \varphi_1 - \varphi_2 + \varphi_3 - \varphi_4 = 0,$

$$[a(\varphi_1)u(\lambda_1)a(\varphi_2)u(\lambda_1)a(\varphi_3)v(\lambda_4)a(\varphi_4)v(\lambda_4) = 1]\Delta_A(\varphi_1,\lambda_1)\Delta_{F_+}(\varphi_2,\varphi_3,\lambda_1,\lambda_4)\Delta_D(\varphi_4,\lambda_4),$$

$$[a(\varphi_1)a(\varphi_2)a(\varphi_3)a(\varphi_4) = 1]\Delta_A(\varphi_1,\lambda_1)\Delta_{F_+}(\varphi_2,\varphi_3,\lambda_1,\lambda_4)\Delta_D(\varphi_4,\lambda_4), \tag{47}$$

(The equations within [ ] mean that both sides of the equation are multiplied by $\Delta\Delta\Delta$.) From eq. (15), we know that the product $AFD \neq 0$ for some value of $\lambda_1$ and $\lambda_4$. So, since there is at least one value of $\lambda_1$ and $\lambda_4$ for which $\Delta_A\Delta_{F_+}\Delta_D = 1$, then the equation in brackets within eq. (47) is valid for all $\varphi_i$ (consistent with $\zeta_+ = 0$,), and we can write

$$a(\varphi_1)a(\varphi_2)a(\varphi_3)a(\varphi_4) = 1, \quad \text{where } \zeta_+ = \varphi_1 - \varphi_2 + \varphi_3 - \varphi_4 = 0, \tag{48}$$

independently of $\lambda_1$ and $\lambda_4$.

Now look at the specific set of angles

$$\varphi_1 = \alpha - \beta, \varphi_2 = \alpha, \varphi_3 = \beta, \varphi_4 = 0,$$
$$a(\alpha - \beta)a(\alpha)a(\beta)a(0) = 1,$$
$$a(\beta - \gamma)a(\beta)a(\gamma)a(0) = 1. \tag{49}$$

The second equation of eq. (49) follows by just relabeling the symbols in the first equation. Multiplying these two equations together gives

$$a(\alpha - \beta)a(\beta - \gamma)a(\alpha)a(\gamma) = 1. \tag{50}$$

If we let β be the average of α and γ in eq. (49) we get

$$\beta = \frac{\alpha + \gamma}{2},$$
$$a(\tfrac{\alpha - \gamma}{2})a(\tfrac{\alpha - \lambda}{2})a(\alpha)a(\gamma) = a(\alpha)a(\gamma) = 1, \tag{51}$$
$$a(\alpha) = a(\gamma).$$

This equation is true because $a^2 = 1$, for any angle. And so, $a$ is a constant, the same for any angle, which is an absurd result. For example, from the second of eqs. (5), when an event occurs, we have

$$A(\varphi_1, \lambda_1)F_+(\varphi_2, \varphi_3, \lambda_1, \lambda_4)D(\varphi_4, \lambda_4) = -1, \quad \zeta_+ = \tfrac{\pi}{2},$$

$$a(\varphi_1)a(\varphi_2)a(\varphi_3)a(\varphi_4) = -1, \tag{52}$$

which cannot happen if all the $a$'s are the same. (We used $\zeta_+ = 0$ in our proofs, but the factorizations, eqs. (18), do not depend on this and are general for all values of $\zeta_{\pm}$.)

And so the whole scheme of assigning hidden variables becomes self-contradictory. We take the result, eq. (51), as a *reductio ad absurbum* refutation of any robust deterministic, realistic theory that attempts to explain Bell-type experiments using the VW state, even with very inefficient detectors.

## 5. Non Perfect Correlations

An immediate extention of the above result for perfect correlations is that the factorization will also be true for experiments at arbitrary angles. We have proven that the functions A, D, and F exist. The only difference, for arbitrary angles, is that $\zeta_k$ can assume any value between $-\pi$ and $+\pi$. However, it is still true that

$$\gamma_k[\varphi_i] = A(\varphi_1, \lambda_1)F_\kappa(\varphi_2, \varphi_3, \lambda_1, \lambda_4)D(\varphi_4, \lambda_4)$$

$$= a(\varphi_1)u(\lambda_1)a(\varphi_2)a(\varphi_3)u(\lambda_1)v(\lambda_4)a(\varphi_4)v(\lambda_4) \tag{53}$$

$$= a(\varphi_1)a(\varphi_2)a(\varphi_3)a(\varphi_4) = +1,$$

since all the $a(\varphi_i)$ are equal. Therefore, from eq. (12),

$$E_{class.}[\varphi_i] = +1. \tag{54}$$

Thus, any experiment at any angle, will give the same result, $+1$. This is a far more stringent result than one gets from the Bell inequalities.

One consequence is that if one raises the objection to our result, that one can never be sure experimentally that one has a perfect correlation, because one cannot measure the angles accurately, it just doesn't matter. Our result is good at all angles. So from an experimental point of view, one only has to show that eq. (54) is not true, regardless of the efficiency of the detectors, and it will disprove any robust model of the system.

## 6. A Single Source *vs*. Two Independent Sources

In the references (6-12) cited earlier, all their reasoning is quantum-mechanical, and the specific question of whether the sources of the down-conversions are two independent lasers, or come from multiple reflections of a single laser, are not important. But in our arguments directly concerning the hidden variables, it is important that they be independent of each other, in other words, that the two down conversions are performed by independent lasers.

We would like to explicitly face the issue here of what differences occur between the two cases, a single source *vs*. two independent sources, for the two sets of particles. If we examine our proof in the 100% efficient case (in paper A), which we gave for two independent sets of particles, and adapt it to the case where both sets of particles come from a single source, we have the condition

$$A(\varphi_1, \lambda) F_{\kappa(\lambda)}(\varphi_2, \varphi_3, \lambda) D(\varphi_4, \lambda) = 1, \quad \zeta_\kappa = 0,$$
$$\zeta_\kappa = \varphi_1 - \varphi_2 + \kappa(\lambda)(\varphi_3 - \varphi_4), \tag{55}$$

where $\lambda$ stands for any particular assignment of hidden variables emanating from the source. From this it follows that

$$A(\varphi_2, \lambda) F_\kappa(\varphi_2, \varphi_3, \lambda) D(\varphi_3, \lambda) = 1, \quad \zeta_\kappa = 0,$$
$$F_\kappa(\varphi_2, \varphi_3, \lambda) = A(\varphi_2, \lambda) D(\varphi_3, \lambda), \tag{56}$$
$$A(\varphi_1, \lambda) A(\varphi_2, \lambda) D(\varphi_3, \lambda) D(\varphi_4, \lambda) = 1,$$

and so the proof will go forward just as in the case with two sources. Thus the same contradiction arises in both cases, and the independence of the two sources is not needed for 100% efficient detectors.

In fact it turns out that one can make a simple model that does satisfy eq. (4) for detectors that are at least 50% efficient, if the four particles, and therefore the $\lambda$'s, are created by a single source, rather than by two independent sources as in the VW experiment. (In this model the counters that detect particles $a$ and $d$ are each 50% efficient, while the BSA is 100% efficient, or other equivalent possibilities.)

However, any of the results that depend on the factorizability of the functions $A$, $D$, and $F$, need two independent sources, as one cannot usefully factor out the $\lambda$ in the one-source equivalent of, say, eq. (23),

$$A(\alpha, \lambda) F(\gamma, \gamma, \lambda) D(\alpha, \lambda) = 1, 0. \tag{55}$$

(For example, one could have $A(\alpha, \lambda) = a(\alpha - \lambda) u(\lambda)$, $D(\alpha, \lambda) = a(\alpha - \lambda) v(\lambda)$, , and $F = AD$, even in the 100% case, since $a^2 = u^2 = v^2 = 1$.)
So all the results for inefficient counters need two independent sources, to allow for factorization. For efficient detectors, a single source is good enough. Of course all of this discussion is meaningless quantum mechanically, but the difference is important if one is trying to model the system deterministically.

## 7. Summary

Quantum mechanically, the singlet state produced by two sources is the same singlet state as that which would be produced by a single source. We believe that because one can reduce the system down to a two-particle state and perform two-particle experiments with it, we can legitimately refer to our results as applying to a two-particle state. (In the standard 2-particle Bohm type EPR experiments, one can think of the decay that produces the singlet state as a "black box", producing this state. Similarly, one can think of the apparatus producing the VW state, the two down-conversions, the BSA and its associated detectors, and all the dials that determine the angles of particles $b$ and $c$, as a black box that creates the entangled state of particles $a$ and $d$. Because the end result is a more complicated state (any one of the Bell states, not just the singlet state), it is a more complicated apparatus, whose output must be monitored, but it is still physically separated from the 2-particle experiment that is performed on particles $a$ and $d$, and so it still plays the role of an isolated black box, defining our 2-particle state.) Thus, the entangled state of the particles $a$ and $d$ are a perfectly normal 2-particle state, but because of the way they were created, they obey our factorization theorem and any realistic, deterministic model of an experiment performed with them will yield much more restricted results than those of the usual Bell-type theorems, and will in fact be inconsistent if they are robust.

To summarize our results, in a two-source experiment using the VW state, any robust deterministic, realistic, local model predicts that every experiment will give the same result, independently of the angles set on the apparatus. flatly contradicting a quantum mechanical calculation. Furthermore, this result is independent of the efficiency of the detectors, and it holds for detectors of any efficiency. (The requirement of robustness limits how low the efficiency can be. Of course at very low efficiencies one has discrimination problems experimentally, when the efficiency is down to the same level as the noise, but that is not our concern here.) So it would seem that deterministic, realistic, local theories are inconsistent with quantum mechanics even in the case of low detection efficiencies, without our having had to resort to any random sampling hypotheses, provided they are robust, which we consider a reasonable limitation.

We would like to thank Prof. David Mermin for several suggestions in the earliest version of this paper, and also Rainer Kaeltenbaek for uncovering a loophole in that version.

**-Figure Caption:**

Figure (1).  <u>Schematic Diagram of the Creation of the Two-Particle State</u>

In this experiment there are two independent down-conversions, one creating the pair of photons $a$-$b$, and the other the pair $c$-$d$.  Each of them undergoes a rotation through the angle $\varphi_i$, and particles $b$ and $c$ enter a Bell-state-analyzer (BSA), which will annihilate them while detecting which Bell state they were in.  If the angles $\varphi_i$ are set properly, as one of the perfect correlation cases, this process forces the particles a and d into a two-particle Bell state.  In the actual experiment the Bell state of $a$ and $d$ is not determined, only their polarizations, but this is sufficient to rule out locally realistic, deterministic theories as an explanation of their observed properties.

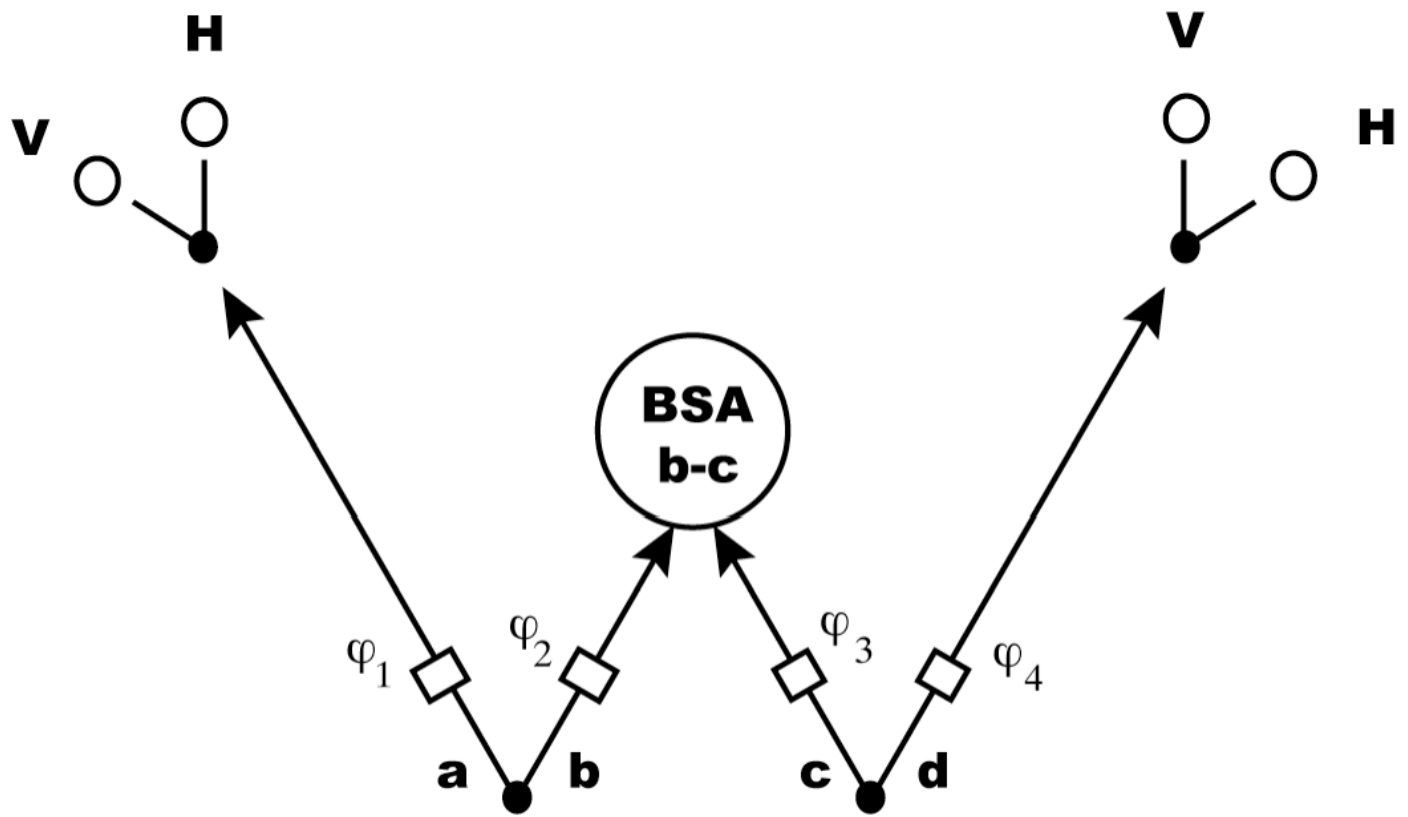

**Fig. (1)**